\documentstyle[preprint,prl,aps]{revtex}
\begin{document}
\draft
\title{Electron-Electron Interactions and
the Hall-Insulator}
\author{Lian Zheng and H. A. Fertig}
\address{Department of Physics and Astronomy,
University of Kentucky, Lexington, Kentucky 40506}
\date{\today}
\maketitle
\begin{abstract}
Using the Kubo formula, we show explicitly that a non-interacting
electron system can not behave like a Hall-insulator,
{\it ie.,} a DC resistivity matrix
$\rho_{xx}\rightarrow\infty$ and $\rho_{xy}=$finite in the zero
temperature limit, as has been observed recently in experiment.
For a strongly interacting electron system
in a magnetic field, we illustrate,
by constructing a specific form
of correlations
between mobile and localized electrons,
that the Hall resistivity
can approximately equal to
its classical value.
A Hall-insulator is realized in this model
when the density of mobile electrons becomes
vanishingly small.
It is shown that in non-interacting electron systems,
the zero-temperature frequency-dependent conductacnce
generally does not give the DC conductance.
\end{abstract}
\narrowtext
A two-dimensional electron system (2DES) in a perpendicular
magnetic field displays
a rich variety of behaviors
at low temperatures as a result of a delicate interplay
between localization and
electron-electron interaction.
The most interesting phenomenon for this 2DES is
the quantum Hall effect (QHE), which has been a focus of
tremendous amount of research effort for more than a decade.
Recently, much attention has been directed to the
behavior of the Hall resistivity for a 2DES in an insulating phase.
Magneto-transport studies of the 2DES have shown\cite{gs1},
among many other
interesting properties, that the Hall resistivity of a
2DES in a perpendicular
magnetic field equals approximately the disorder-free value
$\rho_{xy}\sim B/nec$ for a wide range of applied magnetic field
strength, except
near the QHE plateaus,
where $B$ is the applied magnetic field,
$n$ is electron density of the 2DES, $e$ and $c$ are the electron charge
and speed of light, respectively. The fact that a large portion of the
electrons may become localized by disorder scattering does not
affect the value of $\rho_{xy}$ substantially.
This is in sharp contrast to the behavior of the diagonal resistivity,
which is found to change from $\rho_{xx}\rightarrow0$
in the QHE phase to $\rho_{xy}\rightarrow\infty$ in an insulating phase.
This implies, in particular, the existence of a
new insulating phase with $\rho_{xx}\rightarrow\infty$ and
$\rho_{xy}=$ finite, the so-called Hall-insulating phase, which
has generated much experimental\cite{gs1,exp,jw}
and theoretical\cite{the,the1} work recently.
The goal of this paper is to show that the Hall-insulating behavior
is {\it necessarily}
an interaction effect. It can not be explained with any
independent-particle model.

It is well known that there
exists a different kind of insulator characterized by
$\rho_{xx}\rightarrow\infty$ and $\rho_{xy}\rightarrow\infty$
in a magnetic field, which will be called
conventional insulators here for
convenience. Examples of the conventional insulators are
band insulators in semiconductors and the bulk Anderson insulators.
The difference between a conventional insulator and a Hall-insulator
is determined by the different behaviors in the Hall resistivities
in a magnetic field:
while $\rho_{xy}$
in a conventional insulator
is a measure of density of
{\it mobile} carriers $n_c$; {\it ie.,}
$\rho_{xy}\sim B/n_c$, in a Hall-insulator
it is not, but rather $\rho_{xy}\sim B/n$.
A successful description of the Hall-insulating behavior should
explain the origin of the difference in $\rho_{xy}$.
So far, much of the theoretical work \cite{the}
on the Hall-insulator has been based on calculating
the zero-temperature
frequency-dependent conductivity tensor, where a finite value of
$\rho_{xy}$ is achieved by finding $\sigma_{xx}\sim i\omega$
and $\sigma_{xy}\sim\omega^2$ in the low frequency $\omega\rightarrow0$
limit. We shall see that this is not a complete description.
If one naively takes $\sigma(\omega\rightarrow0)$ to be the DC
conductivity, one would conclude
that all insulators are Hall-insulators. In fact, we will see
that one can not, in general, take the $T\rightarrow0$ limit before
taking $\omega\rightarrow0$.

In an insulating phase, the lowest energy extended state
is above the Fermi level by a finite energy difference.
Low temperature electrical current is carried by electrons
thermally activated to the extended states.
If the activated carriers are effectively decoupled from
the localized electrons in the background, a transport
experiment becomes a measurement of the mobile carriers only.
As we will show below, the system is then a conventional
insulator.
Therefore, in order to
have a Hall resistivity $\rho_{xy}$
which is not a measure of the mobile carrier density $n_c$, but
a measure of the total electron density $n$ as $\rho_{xy}\sim B/n$,
there must exist a strong correlation between the
activated mobile electrons
and the localized electrons in the background.
This simple argument directly implies that a non-interacting
electron system can not become a Hall-insulator.
We will exploit this idea and show that the Hall-insulating behavior
can be realized for electrical conduction by certain kind of
correlated excitation.

There is another unsettled question
related to the Hall-insulator.
It is about its
ground state: is it essentially a pinned
Wigner crystal or a new type of insulator?
The assumption of
a Wigner crystal is found to be qualitatively consistent with
some results from the transport study\cite{thr},
radio-frequency measurement \cite{rf},
and photoluminescence experiment\cite{pl}.
However, a ground state of
disorder-localized electrons rather than a pinned Wigner crystal
has been suggested
by a recently proposed globe phase diagram \cite{gpd}
and the observation of a
similar Hall-insulating phase \cite{jw} at
the Landau level filling factor $\nu>2$.
Our work reported here does not settle the question of which ground state
is correct in a given situation, but rather shows that the
Hall resistivity does not necessarily distinguish between these
possibilities. We will see, instead, that it is the properties
of the excited states that determine the Hall resistivity.

In the following, we will first discuss the non-interacting
2DES and show explicitly that it can not display Hall-insulating behavior.
We will then discuss the
interacting 2DES with correlated excitations
and explain how the Hall-insulating behavior is realized.

We start with the Kubo formula for conductivity $\sigma_{ij}$
\begin{equation}
\sigma_{ij}(T,\omega)=
{i\over\omega}{ne^2\over m}\delta_{ij}+{i\over\omega}\Pi_{ij}(T,\omega).
\label{equ:e1}
\end{equation}

For non-interacting electrons with an arbitrary disorder,
the correlation function $\Pi_{ij}(T,\omega)$
can be evaluated using the
complete set of eigenfunctions of the Hamiltonian. It is straightforward
to obtain the result
\begin{equation}
\Pi_{ij}(T,\omega)={1\over A}\sum_{nm}J_{nm}^iJ_{mn}^j
{f(\epsilon_n)-f(\epsilon_m)\over\epsilon_n-\epsilon_m+
\omega+i0},
\label{equ:pi1}
\end{equation}
where $n$ and $m$ label the eigenstates of the Hamiltonian
($H=\sum_ih_i$ and $h|n>=\epsilon_n|n>$), $J_{nm}$ is the matrix
element of current operator, and $f(\epsilon)$ is the Fermi
distribution function. From the above expression, one can prove the
following properties of the conductivity matrix:

1) Im$[\sigma_{ij}(T,\omega)]$ is an odd function of $\omega$
\begin{equation}
{\rm Im}[\sigma_{ij}(T,\omega)]=O(\omega)+O(\omega^3)+O(\omega^5)+\dots.
\label{equ:e2}
\end{equation}

2) Re$[\sigma_{ij}(T,\omega)]$ is an even function of $\omega$
\begin{equation}
{\rm Re}[\sigma_{ij}(T,\omega)]={\rm Re}[\sigma_{ij}(T,0)]+
O(\omega^2)+O(\omega^4)+\dots.
\label{equ:e3}
\end{equation}

In the following, we show that if
the system is insulating, it must be a conventional insulator.
We will reach
this conclusion by calculating the finite-temperature DC
conductivity and then taking the low temperature limit.

{}From Eq.~(\ref{equ:e2}) and Eq.~(\ref{equ:e3}), we can see that
Im[$\sigma(T)]=0$ for $\omega=0$. We only need to consider the real parts,
which are easily obtained
from Eq.~(\ref{equ:pi1}). The diagonal conductance is
\begin{equation}
\sigma_{xx}(T)\sim\sum_{nm}|J^x_{nm}|^2{\partial f(\epsilon_n)\over\partial
\epsilon_n}\delta(\epsilon_n-\epsilon_m).
\label{equ:cxx1}
\end{equation}

For any localized state\cite{rk1}, we have
\begin{equation}
J_{nm}=<n|J|m>\sim <n|Hr-rH|m>=r_{nm}(\epsilon_n-\epsilon_m).
\label{equ:jnm1}
\end{equation}

Inserting the above expression into Eq.~(\ref{equ:cxx1}), we see that only
extended states can contribute to $\sigma_{xx}$, because
the $\delta$-function requires $\epsilon_n=\epsilon_m$.
Now for an insulator, suppose that the lowest extended state has
an energy $\Delta$ away from the Fermi level, Eq.~(\ref{equ:cxx1})
gives
\begin{equation}
\sigma_{xx}(T)\sim e^{-\beta\Delta},{\rm \ \ \ for \ } T\rightarrow0.
\label{equ:cxx2}
\end{equation}

The off-diagonal conductance has the form
\begin{equation}
\sigma_{xy}(T)\sim\sum_nf(\epsilon_n)\sum_m{J_{nm}^xJ_{mn}^y-
J_{nm}^yJ_{mn}^x\over(\epsilon_n-\epsilon_m)^2}.
\label{equ:cxy1}
\end{equation}

Any localized state can not contribute to $\sigma_{xy}$.
For example, if $<n|$ is localized, then we are allowed to use
Eq.(~\ref{equ:jnm1}), and the contribution from the state
$<n|$ is,
\begin{eqnarray}
&&\sum_m{J_{nm}^xJ_{mn}^y-
J_{nm}^yJ_{mn}^x\over(\epsilon_n-\epsilon_m)^2}
=\sum_m[x_{nm}y_{mn}-
y_{nm}x_{mn}] \\ \nonumber
&&=<n|xy-yx|n>=0.
\label{equ:sm1}
\end{eqnarray}

The summation $\sum_{n}$ in Eq.~(\ref{equ:cxy1})
may then be restricted to extended states
only.
If the lowest energy
extended state is $\Delta$ above the Fermi energy, Eq.~(\ref{equ:cxy1})
gives
\begin{equation}
\sigma_{xy}(T)\sim e^{-\beta\Delta},{\rm \ \ \ for \ } T\rightarrow0.
\label{equ:cxy2}
\end{equation}

Putting together Eq.~(\ref{equ:cxx2}) and Eq.~(\ref{equ:cxy2}), one finds
\begin{eqnarray}
\rho_{xy}(T)&&={\sigma_{xy}(T)\over\sigma_{xx}^2(T)+\sigma_{xy}^2(T)}
\sim{e^{-\beta\Delta }\over e^{-2\beta \Delta}}\\ \nonumber
&&=\infty\ \ \ {\rm for \ \ \ } T\rightarrow0.
\label{equ:rrxy2}
\end{eqnarray}

Similarly, one has $\rho_{xx}\rightarrow\infty$.

The above result shows that the non-interacting electron insulator
is a conventional insulator.
We have not made any approximation in our derivation, except
the restriction that we consider only
non-interacting electrons with a gap between the Fermi level
and the lowest extended state.
We therefore conclude that non-interacting electrons
can not behave like Hall-insulators.

Several authors have described an AC form of the Hall-insulator,
by calculating the zero-temperature
AC conductivity and then considering
the low frequency limit;
{\it ie.}, they took the limit $T\rightarrow0$
first and the limit $\omega\rightarrow0$ second.
  This method
will give a finite value for $\rho_{xy}$ when $\rho_{xx}\rightarrow
\infty$.
It is easy to see that this does not describe the DC conductivity and
could mislead one to conclude that every insulator
is a Hall-insulator.

For an insulator at zero temperature, Re$[\sigma_{ij}(T=0,\omega=0)]
=0$. From Eq.~(\ref{equ:e2}) and Eq.~(\ref{equ:e3}), we have:

\centerline{$\sigma_{xx}^2(0,\omega)+ \sigma_{xy}^2(0,\omega)=O(\omega^2)+\ $
higher
power in $\omega$.}

\centerline{Re$[\sigma_{xy}(T,\omega)]=O(\omega^2)+\ $
higher powers in $\omega$.}

Then we have
\begin{equation}
\varrho_{xy}=
{{\rm Re}[\sigma_{xy}]\over\sigma_{xx}^2+\sigma_{xy}^2}
|_{\omega\rightarrow0}\sim{\omega^2
\over\omega^2}={\rm constant.}
\label{equ:e4}
\end{equation}

If one associates $\varrho_{xy}$ with the
DC Hall resistivity of the system, then
every insulator would be a Hall-insulator.
However, this is not consistent with experiment, suggesting
that interchanging the order of $T\rightarrow0$ and $\omega\rightarrow0$
is in general not valid.

We will now argue that if correlations, in the Laughlin-Jastrow
sense \cite{lau}, are important between conduction
electrons and the localized electrons, then the Hall-insulating
behavior may be obtained.
We consider low temperature activated conduction of a
2D electron gas in an insulating phase.
Let $N=N_c+N_L$ and $n=N/A, n_c=N_c/A, n_L=N_L/A$, where $A$ is
the system size and $N,\ N_c,\ N_L$ are respectively the number
of electron in the system, the number of electrons in
extended states, and the
number of electrons in localized states, with $n,\ n_c,\ n_L$
the corresponding densities.  $N_c$ is activated, so
$N_c\propto e^{-\Delta/kT}$ as $T\rightarrow0$.
We consider only the motion of conduction
electrons and treat the remaining localized electrons as a
scattering source.
Although the electrons below $E_F$ are well localized,
they are still dynamic and responsive to the motion of the conduction
electrons.
This dynamic correlation is a very complicated problem.
We would like to find a effective Hamiltonian to represent
the coupling between the conduction electrons and the localized
electrons, which is much simpler and yet it still gives the
same effect on the motion of the conducting electrons.
This may be achieved using a Chern-Simons approach\cite{zkh}.

The Hamiltonian for the conduction electrons is
\begin{equation}
H_c=\sum_{i=1}^{N_c}[-{i\hbar\over2m}\nabla_i-{e\over c}
{\bf A}^
{ext}({\bf r_i})]^2 +{1\over2}\sum_{ij=1}^{N_c}V({\bf r_i}-{\bf r_j})
+\sum_{i=1}^{N_c}U({\bf r_i})+H_{cL},
\label{equ:ha}
\end{equation}
where ${\bf A}^{ext}$ is the vector potential for applied magnetic
field $B^{ext}$, $V({\bf r})$ is the electron-electron interaction between
a pair of the conducting electrons, $U({\bf r})$ is disorder scattering
potential. $H_{cL}$ is the interaction between the conduction electrons
and the remaining localized electrons, which is important in
strongly correlated systems and presumably is the term responsible for
Hall-insulator behavior. Our next step is to determine a mean-field form of
$H_{c}$ which characterizes
correctly the influence of the localized electrons on the motion of
the conduction electrons.  It is enlightening
to recall an earlier study\cite{cin} on correlated interstitials in
a weakly disordered
Wigner crystal, where one finds that energetically favored
excitations are described by the wavefunction
\begin{equation}
\Psi_{corrl}(z_o)=
\Psi_{uncorr}
\prod_{i=1}^{N_L}(z_i-z_o)^{m_i},
\label{equ:phic}
\end{equation}
where $z_i=x_i-iy_i$ are the lattice electrons in complex notation,
$z_o$ is the interstitial coordinate, and $\nu$ is
the Landau level filling fraction with $B^{ext}=n\phi_o/\nu$.
The values of $m_i$ may be chosen to minimize the energy
of the excitation, and we argued in Ref. [13] that
this may be accomplished if $<m_i>=1/\nu$, where $<>$ denotes
an average over lattice sites. $\Psi_{uncorr}$ describes an
uncorrelated interstitial, which in a Hartree-Fock approximation
would simply be given by a Gaussian orbital at some favorable interstitial
site in the lattice.  The addition of the Jastrow factor
introduces correlations, and it may be shown \cite{cin} that
its introduction converts the excitation into a delocalized
state.  Antisymmetrization corrections between the interstitial
and the lattice electrons have been shown to be small \cite{cin}.

One can clearly see that the physics of such wavefunction
is more general than the Wigner crystal context in which it was derived.
In particular, one can choose $\Psi_{uncorr}$ to be any insulating
state of an excited electron, and introducing the
Jastrow factor creates an excited electron in an extended state, provided
$<m_i>=1/\nu$.

The Jastrow factor in Eq.~(\ref{equ:phic}) may be thought
of as attaching a flux tube of strength $m_i\phi_o$ to the ith
localized electron\cite{lau}. If we consider the longwavelength
response of the conduction electron, then an appropriate mean-field
Hamiltonian will account for interaction with the localized electron by an
additional field\cite{zkh,hlr}$B^{cL}=n_L\phi_o/\nu$, we then have

\begin{equation}
H_c^{MF}=\sum_{i=1}^{N_c}[-{i\hbar\over2m}\nabla_i-{e\over c}
{\bf A}^
{net}({\bf r_i})]^2 +{1\over2}\sum_{ij=1}^{N_c}V({\bf r_i}-{\bf r_j})
+\sum_{i=1}^{N_c}U({\bf r_i}),
\label{equ:ha2}
\end{equation}
where ${\bf A}^{net}$ is the vector potential for the net
magnetic field $\nabla\times{\bf A}^{net}=B^{net}=B^{ext}-B^{cL}
=n_c\phi_o/\nu$. DC transport properties of the conduction
electrons described by $H_c^{MF}$ of Eq.~(\ref{equ:ha2}) are easily
obtained from the Kubo formula in the memory-function
formalism \cite{mem}
\begin{equation}
\rho_{ij}=i\Omega_{ij}+{m\over n_ce^2}\Gamma_{ij},
\label{equ:r1}
\end{equation}
where $i\Omega_{xx}=i\Omega_{yy}=0$ and $i\Omega_{xy}=-i\Omega_{yx}=
B^{net}/(n_cec)=B^{ext}/(nec)$.
Since the electrons are in extended states, the disorder
scattering can be treated perturbatively. To the lowest order
in the disorder potential, $\Gamma_{xy}=\Gamma_{yx}=0$. Denoting
$\Gamma_{xx}=\Gamma_{yy}=1/\tau$, we have
\begin{eqnarray}
\rho_{xx}=&&{m\over n_ce^2\tau}, \\ \nonumber
 \rho_{xy}=&&{B^{ext}\over nec}.
\label{equ:r2}
\end{eqnarray}

The above result shows clearly
that $ \rho_{xy}$ depends only on the external
magnetic field and the number of total electrons in 2D system,
independent of how many of the electrons are localized.

The idea can be recast in
the Drude picture.   Suppose there is a current $j_x=n_cev$
with $\rho_{xx}=1/(n_ce\mu)$. A Hall voltage is generated to
balance the Lorentz force $ \rho_{xy}=E_y/j_x={(vB/c)}{/(n_cev)}=
{B/(n_cec)}$. For uncorrelated conduction electrons, $B=B^{ext}$
and $ \rho_{xy}$ depends on $n_c$, not the total electron number $n$.
One would get $\rho_{xx}\rightarrow\infty$ and $ \rho_{xy}\rightarrow\infty$
when $n_c\rightarrow0$.
For strongly  correlated systems, we have shown that $B=B^{net}
=(n_c/n)B^{ext}$ so that $ \rho_{xy}=B/(nec)$, regardless of the  number
of electrons which are localized.

We have demonstrated
this model of correlation
between the activated conduction electrons and the remaining
localized electrons does make the system a Hall-Insulator.
The key is that we
characterize the interaction between the activated conduction electrons
and remaining localized electrons as flux-tube-like,
and describe them using a Chern-Simons statistical field.
However, we would like to emphasize
that the introduction of a Jastrow factor in
the trial-wavefunction in the
early work\cite{cin} on correlated interstitials
yields extremely low energies with a microscopically realistic
Hamiltonian.
We note also that previous work \cite{gpd} has shown that
a non-divergent $\rho_{xy}$ may be obtained using
the Chern-Simons approach. However, in that case the correlation were
introduced in the ground state rather in the excited states, and
while $\rho_{xy}<\infty$, it was not necessarily equal to $B^{ext}/(nec)$.
By introducing the correlation in the excited states, it is possible to have
localized electrons
for which the flux tube strength are site-dependent,
leading to the classical Hall resistivity.

To summarize, we have shown rigorously with the Kubo formula
that a non-interacting electron systems
can not display the Hall-insulating behavior and
treatments based on zero-temperature frequency-dependent
conductivity
are insufficient to explain this behavior.
Instead, the Hall-insulating behavior should be considered as
an interaction effect.
We have constructed an explicit form for
the strong correlations between the temperature-activated
mobile electrons and the localized electrons in the background
and demonstrated that this kind of correlation
does lead to
the Hall-insulating
behavior.

The authors thank Professor A.H. MacDonald for very
stimulating discussion and for a critical reading of this
manuscript.
This work
is supported by NSF through Grant No. DMR-9202255.


\begin{references}
\bibitem{rk1}
Eq.~(\ref{equ:jnm1})
does not apply to extended states because the formula involves
an integration by parts, for which surface terms have been dropped.
\end{references}
\end{document}